\documentclass{webofc}
\usepackage[varg]{txfonts}   
\begin{document}
\title{Analysis of Galactic molecular cloud polarization maps: a review of the methods}
%
%

\author{\firstname{Fr\'ed\'erick} \lastname{Poidevin}\inst{1,2}\fnsep\thanks{\email{fpoidevin@iac.es}} 
}

\institute{Instituto de Astrofis\'{i}ca de Canarias, 38200 La Laguna,Tenerife, Canary Islands, Spain
\and
           Departamento de Astrof\'{\i}sica, Universidad de La Laguna (ULL), 38206 La Laguna, Tenerife, Spain
          }

\abstract{%
  The Davis-Chandrasekhar-Fermi (DCF) method using the Angular Dispersion
  Function (ADF), the Histogram of Relative Orientations (HROs) and
  the Polarization-Intensity Gradient Relation (P-IGR) are the most
  common tools used to analyse maps of linearly polarized emission
  by thermal dust grains at submilliter wavelengths in molecular
  clouds and star-forming regions. A short review of these methods is
  given. The combination of these methods will provide
valuable tools to shed light on the impact of the magnetic fields on
the formation and evolution of subparsec scale hub-filaments that will be
mapped with the NIKA2 camera and future experiments. 
}
\maketitle
\section{Introduction}
\label{intro}

\begin{figure}[h]
\begin{center}
  \includegraphics[scale=3.3]{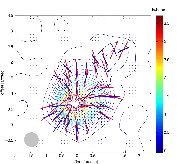} 
  \includegraphics[scale=0.14,angle=0]{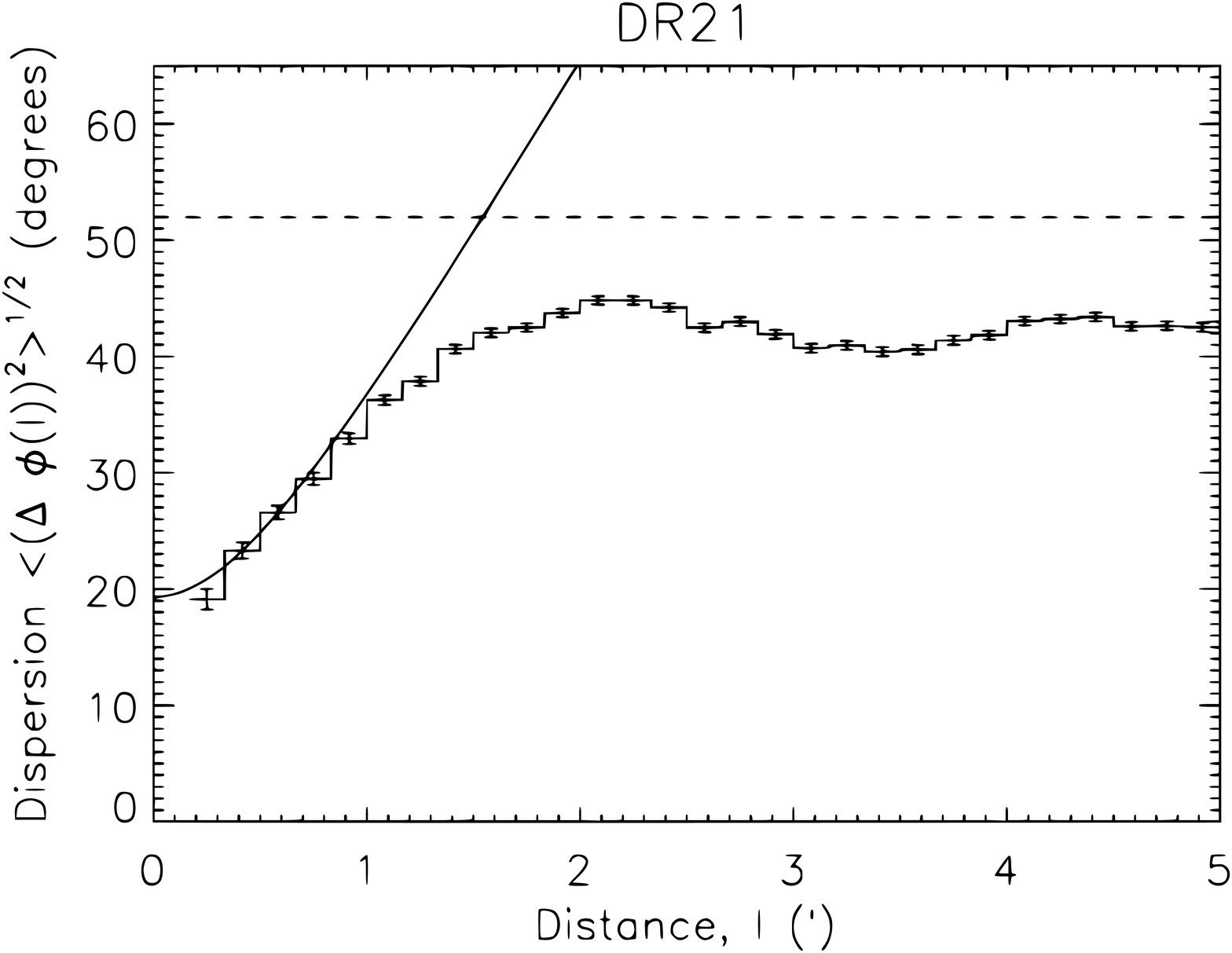} 
\label{fig-1}       
\caption{Left: Figure 1 from \cite{koch2012} showing
  the Submillimeter Array (SMA) 870 $\mu$m polarization map of
  the Collapsing core W51 e2 obtained by \cite{tang2009}. The thick red segments show the magnetic
  field orientation after rotation by 90$^\circ$ of the
  polarization pseudo-vectors (see Introduction)
  where polarization was detected such that $p/\sigma_{p}>3$.
  The degree of polarization is not represented in this map.
  The blue pseudo-vectors show the gradient directions
  (see Section~\ref{sec-2} and Section~\ref{sec-3}) of the dust
  emission continuum mapped at 1.3 mm by \cite{lai2001}
  with the Berkeley-Illinois-Maryland Association (BIMA). 
  Right: Figure 2 from \cite{poidevin2013} showing the ADF of the SCUBA
JCMT DR21 850 $\mu$m polarization maps (see Section~\ref{sec-1}).}
\end{center}
\end{figure}

This article gives a short review of the main methods used to analyse
linear polarization maps. Given the more and more prolific literature
on the subject, not all the articles using each technique are cited but
the generic articles introducing the conceptual ideas behind each
procedure are used as references. Additional methods employing wavelet
transform analysis technics and Bayesian inference statistical
tools currently in development are not discussed here, and we refer
the interested reader to the method investigated by \cite{robitaille2019}
and to the IMAGINE consortium project \cite{imagine2018}, respectively.
The review focuses on the analysis of maps
obtained from the observation of linearly polarized thermal dust
emission at submillimeter (submm) wavelengths toward Galactic Molecular
Clouds (MCs) and protostellar cores, but the same methods can be used on maps
obtained from simulations.
Independently of the dust grain alignment mechanism that is considered
(see \cite{and2015} for a review on this subject), the main accepted
current picture is that of dust grains aligned perpendicular to
the local magnetic field direction pervading the Interstellar Medium
(ISM). Each polarization pseudo-vector displayed in one pixel of a polarization map
is therefore an average measurement of the weighted contribution by all
dust grains along a given Line-Of-Sight (LOS) in a direction
perpendicular to the average magnetic field on the Plane-Of-Sky (POS).
From the measurements of the Stokes parameters $I$, $Q$ and $U$
the total fraction of polarization ($p=\frac{\sqrt{(Q^2+U^2)}}{I}$) is
often represented by the length of the pseudo-vector and the polarization angle 
(P.A.; $\theta=\frac{1}{2} \times \rm arctan (\frac{U}{Q})$) by
its orientation with respect to a given reference frame.
Other representations of $p$ exists in the literature and maps
showing only drapery patterns of the POS magnetic field lines,
or of the P.A.s, are more and more common.
An example of submm polarization is shown in figure~\ref{fig-1} (left panel).

\section{Davis-Chandrasekhar and Fermi (DCF) method and Angular Dispersion
  function (ADF)}
\label{sec-1}

The DCF method was introduced in the middle of the twentieth century by
\cite{davis1951} and \cite{cf1953}. It was first designed to get
estimates of the POS magnetic field strength, $B_{pos}$, assuming a turbulent
diffuse ISM. It was applied on polarimetry data
obtained on star fields by \cite{hiltner1951} at visible wavelengths.
In this regime, the polarization is produced by dichroic absorption of
starlight by dust grain layers pervading the diffuse ISM, in
a direction perpendicular to the one produced in
emission at submm wavelengths by identical polarizing
dust grains (see \cite{planck_ir_xxi}). Inferred from the
data is a large-scale uniform magnetic field along the Galactic Plane
(GP). The fluctuations around the mean of the distribution of the polarization
pseudo-vectors are assumed by \cite{davis1951} and \cite{cf1953}
to be produced by Alfv\'en waves that are coupled
to the gas such that there is equipartition
between kinetic and perturbed magnetic energies.
With this model $B_{pos}$ is a function of the
ISM gas density $\rho$, of the gas velocity dispersion $\sigma_{v}$, and of the
polarization angle dispersion $\sigma_{\theta}$, such that:
$B_{pos} = Q \sqrt{4 \pi \rho} \frac{\sigma_{v}}{\sigma_{\theta}}$,
where $Q$ is a factor of proportionality.  With this method,
\cite{cf1953} calculated a diffuse ISM $B_{pos}$ estimate of a few $\mu$G consistent
with those obtained with other independent methods. Later on, in the 1990s, when polarimetry
detectors at submm wavelengths started to be sensitive enough,
the DCF method was used on submm maps of molecular clouds,
meaning transposed to spatial scales 1000 to 10000 times smaller
than the GP scale in regions of density two orders or more magnitudes that of
the diffuse ISM density. The first comprehensive study
was done by \cite{gonatas1990} from 10 measurements obtained at 100$\,\mu$m
with the Kuiper Airborne Observatory (KAO) in the Orion Nebula, leading
to $B_{pos}$ estimates of a few mG. The DCF method has been tested
numerically by \cite{ostriker2001} and \cite{dfg2008} and some
refinements to the calculations of $B_{pos}$ have been proposed. A review on
the values of $Q$ has been given by \cite{poidevin2013}. 

\textbf{Improvements to the Method}:
Further major improvements to the method were designed
by \cite{hildebrand2009}, \cite{houde2009} and \cite{houde2011}.
The Angular Dispersion Function (ADF) was introduced by \cite{hildebrand2009} to
avoid inaccurate estimates of magnetohydrodynamic or turbulent
dispersion, as well as to avoid inaccurate estimates of $B_{pos}$,
due to the large-scale, non turbulent field structure.
The ADF is expressed by
$<\Delta \Phi^2(l)>^{1/2} \equiv \left\{ \frac{1}{N(l)}
  \Sigma_{i=1}^{N(l)}[\Phi(x) - \Phi(x+l)]^2    \right\}^{1/2}$,
where $\Phi(x)$ is the angle asssociated to the projected POS magnetic
field vector $B(x)$ at position $x$ in a map. The difference in angle
between two points is obtained by $\Delta \Phi(l) \equiv \Phi(x) -
\Phi(x+l)$, and is calculated between the $N(l)$ pairs of vectors
separated by displacement, or lag, $l$. $<...>$ denotes an average and
$l=|l|$. The square of the ADF, a second order structure
function, is also often used (e.g. \cite{dfg2008}). One example of ADF
is shown in Figure~\ref{fig-1} (right panel), where the angular
dispersion $b$ is the intercept of the fit at $l=0$.
Correlations in
polarization angles at lags $l$ smaller than the telescope
beam ($1.22 \lambda / D$) or than the turbulent correlation length
($\delta$) have to be avoided.
The fit is ideally applied on the set of points calculated by
taking into account the measurement uncertainties and such that
the lag distance $l$ is smaller that the typical length scale $d$ for
variations in the large-scale magnetic fields. Once $b$ is estimated
from a map this method also provides an estimate of the turbulent
to large-scale magnetic field strength ratio such that
$\frac{<B_{t}^{2}>^{1/2}}{B_{pos}}=\frac{b}{\sqrt[]{2-b^2}}$,
and the POS strength of the large-scale component is estimated by
$B_{pos} \simeq \sqrt[]{8 \pi \rho} \frac{\sigma_{v}}{b}$.
The method to take into account the effect of the signal integration
through the thickness of the clouds as well as across the area
sustended by the telescope is fully incorporated by   
\cite{houde2009}. The authors also show how to evaluate the turbulent
magnetic field correlation length scale from polarization maps obtained
with sufficiently high resolution and high enough spatial sampling
rate. Further examples are given and discussed by \cite{houde2011}
as well as the application of the technique to interferometry measurements.

\begin{figure}[h]
\begin{center}
  \includegraphics[scale=0.38,angle=90]{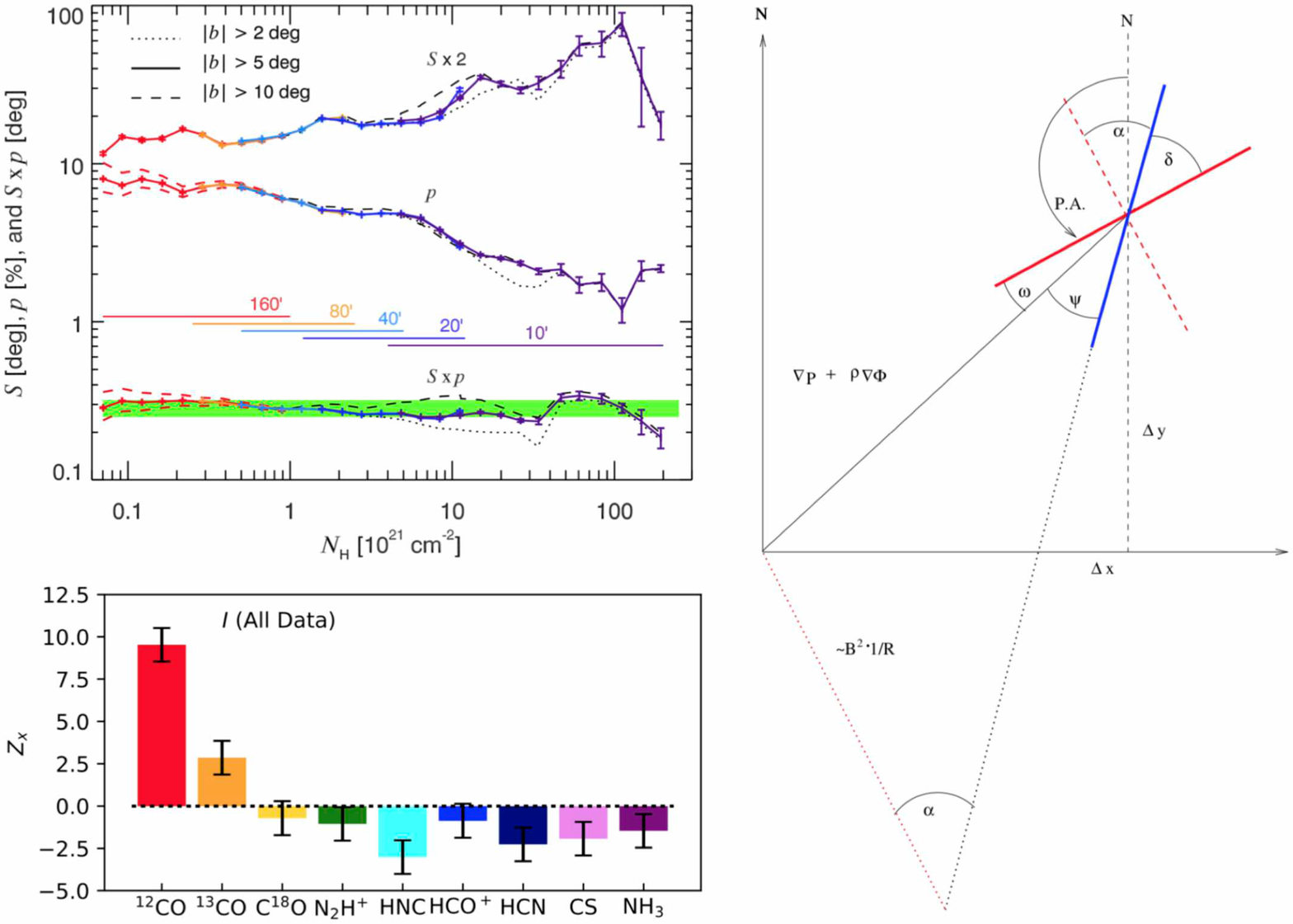}
\label{fig-2}       
\caption{
  Top-Left: Figure 11 from \cite{planck2018_xii} showing the
$S \times p$ relation as a function of column density $N_{\rm H}$ where 
 $w$ is the map resolution parameter (see Section~\ref{sec-1}).
 Bottom-Left: Figure 6 from \cite{fissel2019} showing the PRS $Z_{x}$
 corrected for oversampling obtained with several molecular tracers (see Section~\ref{sec-2}).
 Right: Figure 3 from \cite{koch2012} showing the 90$^{\circ}$ rotated
 P.A. $\alpha$, and the angle associated to the gradient orientation $\psi$, used to
 retrieve information on the magnetic field strength and significance (see Section~\ref{sec-3}).   
}
\end{center}
\end{figure}

\textbf{Results:}
The DCF method has been applied to data from many Galactic
MCs obtained on sky patches, including Gould belt MCs  (e.g.
\cite{coude2019}) and some of the
closests low, intermediate or high star-forming regions.
Estimates of $B_{pos}$ lie typically in the range of a few $\mu$G
to a few mG. In OMC-1 the turbulent correlation length is
estimated to $\delta \approx $ 10 mpc (e.g. \cite{houde2009}).
Independently of the uncertainties coming from the
propagation of the errors, the estimates of $B_{pos}$ and $\delta$
will rely on the choice (or availabilty) of the gas
tracers and of the value of $Q$ used for the calculation of $B_{pos}$
(e.g. see discussion in \cite{pattle2017}). In
principle accurate and reliable results can be obtained with polarization data of
sufficient spatial resolution and high enough spatial sampling rate
(\cite{houde2009}).   
The DCF method has been applied to a large fraction of the sky by \cite{planck_ir_xix}
and on the full sky by \cite{planck2018_xii}. Using the Planck Release
3 \cite{planck2018_xii} have obtained the following relation between
the ADF, $S$, and the fraction of polarization, $p$, as a function of the
map resolution, $w$:  
$<S_{p}>=\frac{0^{\circ}.31}{p} \left( \frac{w}{160^{'}} \right)$.
The results are displayed in Figure~\ref{fig-2} (top-left panel), and show that down to
a resolution of $10'$ the systematic decrease in $p$ with $N_{\rm H}$ is
determined mostly by the magnetic-field structure. 
At a lower resolution of $2.5'$, using the $500 \mu$m BLASTPol
polarization map of Vela C, \cite{fissel2016} discuss the dependence
of $p$ on the dust temperature and on $N_{\rm H}$ and show that
$p \varpropto N_{\rm H}^{-0.45} S^{-0.60}$,
suggesting that dust grain alignment properties may also contribute to the
decrease of $p$ in some conditions.

\section{Histogram of Relative Orientations (HROs) between ISM tracer structures and magnetic field structures}
\label{sec-2}

This method has been designed by \cite{soler2013} and subsequently
applied to many molecular cloud regions (e.g. \cite{soler2017}, \cite{planck_ir_xxxv}).
The concept of the method relies on the calculation of the gradient of
the column density structure provided by a given material tracer 
(e.g. $N_{\rm H}$) in the pixels of a polarization map. As an illustration
the blue pseudo-vectors displayed in Figure~\ref{fig-1} (left panel), show the
gradient orientations obtained from a dust emission continuum map.
Once the gradients are quantified their relative orientation with
the POS magnetic field orientations inferred from the P.A.s can be
estimated and the HROs can be analysed. Several methods exist to
calculate the gradients in a map as a function of the morphology structure
of the object one is interested to isolate (see e.g. \cite{alina2019} and references
therein). HROs can be built by considering the relative
orientation of P.A.s with respect to the average orientation of a
large-scale structure identified in a map (e.g. \cite{alina2019}).
More frequently, HROs are calculated by considering
the relative orientation angle $\phi$ between the POS magnetic field
$<\widehat{B}_{pos}>$ and a line tangent to the local iso-contour
(see \cite{soler2013}, \cite{soler2017}, \cite{fissel2019})
which is equivalent to the angle between
the polarization direction $\widehat{E}$ and the intensity gradient $\bigtriangledown I$:
$\phi = \rm arctan (| \bigtriangledown I \times \widehat{E} |, \bigtriangledown I. \widehat{E})$. 
Statistical measures of HROs have been quantified using
the histogram shape parameter (e.g. \cite{soler2013}, \cite{soler2017}),
$\zeta=\frac{A_{0}-A_{90}}{A_{0}+A_{90}}$, where $A_{0}$ is a measure
of the total number of points in the quartile $0^{\circ} < \phi <
22^{\circ}.5$, and $A_{90}$ a measure of the same quantity in the
quartile $67^{\circ}.5 < \phi < 90^{\circ}$. Smoother and more
accurate definitions of $\zeta$ have been investigated by
\cite{jow2018}. For this Rayleigh statistics $Z$ is used to test
whether a given set ${\theta_{i}}$ of $n$ independent angles
are uniformly distributed within the range$[0, 2\pi]$.
Using the relation $\theta = 2 \phi$, where $\phi$ is the relative
orientation angle discussed above, 
\cite{jow2018}
showed that the Projected Rayleigh Statistic (PRS) of $Z$:
 $ Z_{x} = \Sigma_{i}^{n_{\rm ind}} \rm cos
    \theta{_i}/\sqrt{n_{\rm ind}/2}$,
where $n_{\rm ind}$ is the number of independent data samples in the map, 
can be used to test for a preference for perpendicular or parallel
alignment between the magnetic field orientations and the iso-contours.
The variables $\zeta$ or $Z_{x}$ are often calculated on samples of
data lying in different ranges of intensity of the material tracer
map, and used to explore variations of the relative orientations
between the magnetic field structures and the ISM morphology structures
as a function of the column density or number density parameters.
Figure~\ref{fig-2} (bottom-left panel) shows the
PRS obtained by \cite{fissel2019} using several
molecular tracers. For each tracer, $Z_{x} >0 (<0)$ indicates the $I$ structure
preferentially aligns parallel (perpendicular) to $<\widehat{B}_{pos}>$.

\textbf{Results:} The main trend found from the analysis of HROs derived
with lines tangent to local iso-contours is that
the POS magnetic field orientations change from mostly parallel (or
not clearly defined) to perpendicular 
with respect to the molecular cloud structures probed with
a given material tracer (\cite{soler2017}, \cite{planck_ir_xxxv},
\cite{fissel2019}). This translates by a change of orientation from
lower to higher column densities $N_{\rm H}$. In Vela C the transition
is estimated to occur at molecular hydrogen number density of
approximately $n_{\rm H} \approx 10^{3} \rm cm^{-3}$
(\cite{fissel2019}). This local iso-contours approach has been
tested by simulations (e.g. \cite{soler2013}).  
In their study \cite{alina2019} first produce a component separation
analysis of the magnetic fields in the diffuse ISM
and in higher column density regions,
and use image analysis technics to extract filaments
and clumps embedded in different background column densities, i.e.
showing density contrast varying with their environment. Their
analysis of the HROs obtained between filaments, embedded
clumps and internal and background magnetic field orientations
lead to a more complex picture than in other studies. Overall
their results support the possibility of magnetic fields strong enough
to influence the formation of molecular clouds and also of their embedded clumps.

\section{Polarization-Intensity Gradient Relation (P-IGR)}
\label{sec-3}

This method has been designed by \cite{koch2012} to study star-forming
regions. In the case of negligible viscosity and infinite conductivity (ideal
MHD case) the force equation is given by the following expression
$\rho \left( \frac{\partial }{\partial t} + \mathbf{v} . \bigtriangledown  \right)
\mathbf{v} = - \bigtriangledown \left( P + \frac{B^{2}}{8 \pi} \right)
- \rho \bigtriangledown \phi + \frac{1}{4 \pi} (\mathbf{B} . \bigtriangledown
) \mathbf{B}, $
where $\rho$ and $\textbf{v}$ are the dust density and velocity,
respectively. $\textbf{B}$ is the magnetic field, $P$ is the
hydrostatic dust pressure, $\phi$ is the gravitational potential
resulting from the total mass contained in the region of interest and,
$\bigtriangledown$ denotes the gradient. The left-hand side term in the
equation represents the resulting motion of the dust produced by
the right-hand side terms which are the gradients of the hydrostatic
pressure terms of the gas, the magnetic field, and the gravitational
potential together as well as the magnetic field tension term (last term).
The force equation can be transformed and under several assumptions
the magnetic field strength can in principle be derived geometrically
at each position of a polarization map and expressed by: 
$B = \sqrt{\frac{\rm sin(\psi)}{\rm sin(\alpha)} ( \bigtriangledown P +
  \rho \bigtriangledown \phi ) 4 \pi R}$,
where $\alpha$ = P.A.-90$^{\circ}$, and $\psi$ is the angle associated to the
gradient orientation. Figure~\ref{fig-2} (right panel), illustrates the 
terms displayed in this equation. The red and blue pseudo-vectors
displayed in the figure can be compared to those displayed in
figure~\ref{fig-1} (left panel) where the gradients in intensity are estimated
assuming central symmetry towards the brightest pixel in the map, but
the method can be generalized to arbitrary cloud shapes. 
An important outcome of the method  is given by the magnetic field significance:   
$\Sigma_{B} \equiv \left( \frac{\rm sin(\psi)}{\rm sin(\alpha)}\right)_{local} = \left(
\frac{F_{B}}{\mid  F_{G}+F_{P}  \mid} \right)_{local}$, 
which gives a direct physical meaning to the factor $\frac{\rm
  sin(\psi)}{\rm sin(\alpha)}$ derived geometrically from a map.

\textbf{Results:} The method provides a quantification of
the local significance of the magnetic field force compared
to the other forces in a model-independent way. In W51 e2 it allows
derivations of the azimuthally averaged radial profile $B (r ) \sim
r^{-1/2}$ (\cite{koch2012}). The potential of the method is explored in
additional works (e.g. \cite{koch2018} and references 
therein). In their study, \cite{tang2019}
propose that in G34 the varying relative importance between magnetic field, gravity,
and turbulence from large-to-small-scale drives and explains the different
fragmentation types seen at subparsec scales
(no fragmentation, aligned fragmentation and clustered fragmentation).

\section{Conclusions and perspectives}
\label{conclusions}

The methods discussed above are complementary to each other. 
If submm polarization maps are available at different resolutions they
are suited to explore the role of magnetic fields,
turbulence and gravity on different spatial scales. Their combination
starts to give insights on the interplay between magnetic fields,
gravity and turbulence (e.g. \cite{tang2019}) and they are promising tools to shed light
on the physics of hubs-filaments (e.g. \cite{andre2019})
detected at subparsec scales in molecular clouds.
In addition to these methods, getting estimates of the mean
inclination with respect to the LOS of the large-scale
ordered magnetic field considered as the main factor
regulating the mean level of polarization in a map can help guiding
the overall interpretation.
In this regard, a method first proposed by \cite{poidevin2013} 
has been further explored by other authors (e.g. \cite{chen2019}).
Multi-wavelengths submm polarimetry of a region can give
insights about polarization dust grain properties and also help to put constraints
on the interpretation of the maps.
On this matter we refer the reader to 
\cite{and2015}, \cite{vaillancourt2012}, \cite{shariff2019} and
references therein.

\textbf{}

\textbf{Acknowledgments:} FP acknowledges
support from the Spanish Ministerio de Economia
y Competitividad (MINECO) under grant numbers
ESP2015-65597-C4-4-R and ESP2017-86852-C4-2-R.

%
%
%

\end{document}